\documentclass[twocolumn,showpacs,preprintnumbers,amsmath,amssymb]{revtex4}
\usepackage{graphicx}
\begin{document}
\title{Scattering wave function approach to multi-terminal mesoscopic system with spin-orbit coupling}
\author{Yongjin Jiang}
 \affiliation{Department of Physics,
ZheJiang Normal University, Jinhua, Zhejiang, 321004,P.R.China}
\author{ Liangbin Hu}
\affiliation{Department of physics and Laboratory of photonic
information technology, South China Normal University, Guangdong
510631, P. R. China }

\begin{abstract}
In this paper,we present a detailed formulation to solve the
scattering wave function for a multi-terminal mesoscopic system
with spin-orbit coupling. In addition to terminal currents,  all
\emph{local} quantities can be calculated explicitly by taking
proper ensemble average in the Landauer-Buttiker's spirit using
the scattering wave functions. Based on this formulation, we
derive some rigorous results for equilibrium state. Furthermore,
some new symmetry relations are found for the typical two terminal
structure in which a semiconductor bar with Rashba or/and
Dresselhaus SO coupling is sandwiched symmetrically between two
leads. These symmetry property can provide accuracy tests for
experimental measurements and numerical calculations.
\end{abstract}

\maketitle
\section{Introduction}
During the past two decades, there is a fundamental progress in
the understanding of transport properties in mesoscopic
systems\cite{data}. Quantum coherence in these systems plays a
much more important role than in macroscopic dissipative systems.
Accordingly, traditional approaches based on quasi-classical
picture (e.g., Boltzmann approach) of carriers ceased to work in
these systems. Fortunately, a well known new approach was built up
to take into account quantum coherence adequately, i.e.,
Landuer-Buttiker theory(LBT)\cite{data}. In this theory, a
mesoscopic system under consideration is connected to electric
contacts through some ideal leads and the transport properties are
determined by carrier scattering probability between these leads.
In practice, the central part of LBT is the quantum mechanical
\emph{scattering problem} for a particle incident from each lead.
However, under most circumstances, only the terminal quantities
are actually concerned (most real experiments probe only terminal
currents instead of current density, or any other local quantities
in the system). Then, all one need is the scattering matrix
between the leads and the terminal conductances can be simply cast
into a beautiful formula expressed by Green's
functions\cite{data}. This formula is frequently used in
literature. Up to our knowledge, no one has bothered himself
previously to solve the whole \emph{scattering wave function}
explicitly. However, as is easy to see, the scattering wave
functions are needed, if we want to calculate \emph{local }
properties in the Landauer-Buttiker Scenario.

 Recently, there is a hot topic in the field of
spintronics, called spin hall effect(SHE)\cite{
Hirsch1999,MurakamiScience2003,SinovaPRL2004}. SHE refers to the
phenomena that when  a longitudinal electric field (or charge
current) is applied in a semiconductor strip, transverse spin
current and/or ensuing spin polarization near transverse
boundaries can be induced. Such an effect opens a possible new way
to manipulate spin degree of freedom by all-electrical means,
which is a big goal to the frontier research community in
semiconductor industry\cite{Dassarma2004,semibook2002}. Some years
ago, two theoretical works predicted such an
effect\cite{MurakamiScience2003,SinovaPRL2004} as an intrinsic
property of spin-orbit coupled semiconductor band structure. Since
then, a lot of theoretical efforts are made to clarify some
fundamental controversies in this
field\cite{JPHu2003,RashbaPRB2003,InouePRB2004,MishchenkoPRL2004,SQShen2004,NikolicPRL2005,NikolicPRB2005,NikolicPRB2006,LShengPRL2005,JLi2005,Mlushkov05,HalprinPRL,SarmaPRL,Usaj2005,Sarma06,Ma2004,shijunren,Jsinova2005,Zyang2005,YJJiang2005}.
There are now also two central experimental results reporting the
\emph{local} measurement of opposite spin polarization near two
transverse boundaries\cite{KatoScience2004,WunderlichPRL2005},
which have attracted many theoretical interpretations. Despite
many important progresses and consensus made, there sill remain
some basic difficulties\cite{Jsinova2005}.

Different theoretical approaches for spin dependent transport are
employed in this field. Among them, the Landauer-Buttiker
formalism is suitable to address the ballistic transport regime
and has the merit of fully taking into account the phase
information. This approach is in some sense mutually complementary
to Boltzmann approaches based on the quasi-classical wave packet
picture. Based on this approach, some important numerical results
has been achieved in the topic of intrinsic
SHE\cite{NikolicPRL2005,NikolicPRB2005,NikolicPRB2006,LShengPRL2005,JLi2005,Usaj2005,Zyang2005,YJJiang2005}.
However, it is worthy to point out, that most previous works based
on the Landauer-Buttiker's picture does not solve the scattering
wave functions
explicitly\cite{NikolicPRL2005,NikolicPRB2005,LShengPRL2005,JLi2005,Usaj2005}.
The normal green's function formula\cite{data} compute terminal
conductances but not \emph{local} quantities such as spin density.
On the other hand, the so called Landauer-Keldysh
formalism\cite{NikolicPRL2005,NikolicPRB2005,Usaj2005} used by
some authors which can compute local quantities is \emph{not}
manifestly single particle description and lacks intuitional
simplicity. So, in order to calculate the non-equilibrium local
quantities in a transparent manner, it is appealing to solve the
single particle scattering wave function in the Landauer-Buttiker
geometry explicitly.

In this paper, we will set up the linear equations to solve the
scattering wave function in the multi-terminal Landauer geometry
explicitly. With single particle wave function at hand,  we can
calculate any non-equilibrium \emph{local} quantities(e.g.,spin
density), by taking proper ensemble average in the
Landuer-Buttiker spirit. Furthermore, we will derive some rigorous
results in the scattering wave function description. These
include: (1)deduction of the green's function formula for terminal
spin current widely used in literature, (2) rigorous proof of some
important properties for equilibrium state in the Landauer's
structure, (3) revelation of some new symmetry relations in a
typical two terminal structure. These symmetry property  can
provide accuracy tests for experimental measurements and numerical
calculations.

\section{\label{sec:1}Description of the method to solve scattering wave function}

\begin{figure}[tbh]
\includegraphics[width=7cm,height=6cm]{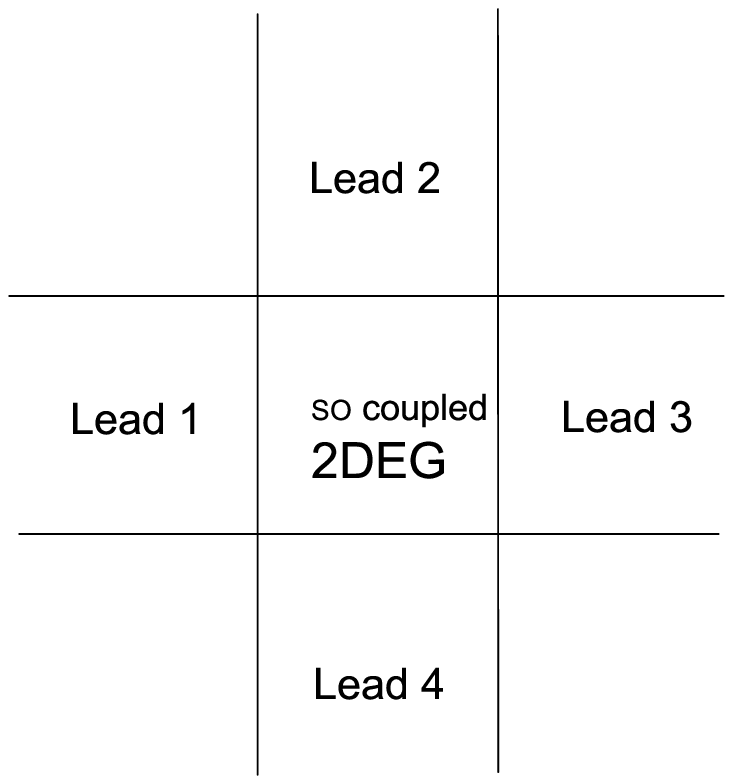}
\caption{schematic geometry of multi-terminal structure }
\end{figure}

We will consider a general mesoscopic structure in which a
spin-orbit coupling system is attached to several ideal leads, as
shown in Fig.1. The p'th lead is extending to reservoir with
chemical potential $\mu_{p}$ at infinity. In a discrete
representation, both the SO system and leads  are modelled by a
nearest neighbor tight binding(TB) Hamiltonian. We assume a
tunable coupling between p'th lead and SO system by hopping
interaction $t_{ps}$. The Hamiltonian for the total system reads:
\begin{equation}
{\cal H}=H_{leads}+H_{sys}+H_{int} \label{eq:1}
\end{equation}
Here, $H_{leads}=-\sum_{p}t_{p}\sum_{<p_{i},p_{j}>\sigma}
(C_{p_{i}\sigma}^{\dag}C_{p_{j}\sigma }+h.c.)$ is the TB
Hamiltonian for the half infinite leads and $C_{p_{j}\sigma}$
denotes annihilation operator for j'th site with spin index
$\sigma$ on p'th lead.
$H_{int}=-\sum_{p}t_{ps}\sum_{n}(C_{p_{n}\sigma}^{\dag}C_{R_{n}\sigma
}+h.c.) $ describes coupling between SO system and leads, where
$t_{p's}$ is the tunable nearest-neighbor hopping parameter
between the lead $p'$ and the SO system. ($p_{n},R_{n}$) is the
nearest pair across the lead/system interface.
$H_{sys}=H_{0}+H_{so}$ is the Hamiltonian for the SO coupling
system in which $H_{0}=-t\sum_{<R_{i},R_{j}>\sigma}
(C_{R_{i}\sigma}^{\dag}C_{R_{j}\sigma }+h.c.)$ is the free band
Hamiltonian,where $C_{R_{j}\sigma}$ is annihilation operator in
the SO system,and $H_{so}$ denotes SO coupling term. For a typical
Rashba model, we write:
\begin{equation}
H_{so}^R=-t^{R}\sum_{R_{i}}[i(\Psi _{R_{i}}^{\dag }\sigma ^{x}\Psi
_{R_{i}+y}-\Psi _{R_{i}}^{\dag }\sigma ^{y}\Psi _{R_{i}+x})+h.c.]
\label{eq:rashba}
\end{equation}
Where $t^{R}$ is the Rashba coupling strength and $\Psi
_{R_{i}}=(C_{R_{i},\uparrow },C_{R_{i},\downarrow })$ is the
vector representation of spinor annihilation operators.

In the following, we will adopt local coordinate frame. Generally
we use double coordinate index $(x_{p},y_{p}),x_{p}=1,...\infty$
and $y_{p}=1,...,N_{p}$ for sites in lead p, and $(x_{s},y_{s})$
for sites in 2DEG system.We will also use single coordinate $y$ or
$y_s$ to represent one site on the leads or in the system when
there is no risk of confusion. Furthermore, we denote $n_{py}$ as
the boundary site in the system near to \emph{first row} site
$(x_{p}=1,y_{p})$ in lead p.

Now, suppose there is an incident wave $e^{-ik_{m}^{p}x_{p}}\chi
_{m\sigma }^{p}(y_{p})$ in lead $p$ (we assume the m'th transverse
mode in lead $p$ is conducting for incident energy $E$), where
$k_{m}^{p}$ is the longitudinal wave vector, $\chi _{m}^{p}$ is
the wave function of m'th transverse mode and $\sigma$ is the spin
index of the incident carrier. Hereafter we choose z axis(normal
to the 2DEG plane) as the quantization axis
for spin. Generally, the  wave function in lead $p'$ can be written as:%
\begin{eqnarray}
\psi _{\sigma \prime }^{pm\sigma}(x_{p'},y_{p'})=\delta
_{pp'}\delta _{\sigma \sigma \prime }e^{-ik_{m}^{p}x_{p}}\chi
_{m\sigma }^{p}(y_{p})\nonumber\\ +\sum_{m'\in {p'}}\phi
_{p'm'\sigma ^{\prime }}^{pm\sigma }e^{ik_{m'}^{p'}x_{p'}}\chi
_{m'\sigma ^{\prime }}^{p'}(y_{p'}) \label{eq:swavefunction}
\end{eqnarray}
where $\phi _{p'm'\sigma ^{\prime }}^{pm\sigma }$ is the
scattering amplitude from in-going mode $(m\sigma)$ in lead $p$ to
out-going mode $(m'\sigma')$ in lead $p'$. The longitudinal wave
vectors is determined by:
\begin{eqnarray}
k_{m'}^{p'}&=&\cos^{-1}(\frac{E-\varepsilon_{m'}^{p'}}{-2t_{p'}}),
\left\vert {E-\varepsilon_{m'}^{p'}}\right\vert \leq {2t_{p'}} \nonumber\\
k_{m'}^{p'}&=&i\cosh^{-1}(\frac{E-\varepsilon_{m'}^{p'}}{2t_{p'}})+\pi,
{E-\varepsilon_{m'}^{p'}}> {2t_{p'}}\nonumber\\
k_{m'}^{p'}&=&i\cosh^{-1}(\frac{E-\varepsilon_{m'}^{p'}}{-2t_{p'}}),
{E-\varepsilon_{m'}^{p'}}<{-2t_{p'}}\nonumber\\
\label{eq:wave vector}
\end{eqnarray}

In the Eq.(\ref{eq:wave vector}), the first line describes real
$k_{m'}^{p'}$ and the corresponding mode is conducting. The last
two lines have imaginary $k_{m'}^{p'}$ value and the corresponding
mode is non-conducting, or, \emph{evanescent}. The evanescent mode
describes local components of scattering wave function in leads
which decay exponentially away from the lead/system interface. It
contributes zero to terminal current. The incident energy $E$ lies
out of the band continuum of such modes.

Now let's use a column vector $\psi_{s}$ of length $2N$ ($N$ is
the number of lattice sites in the SO system) to represent the
wave function in the SO system $\psi
_{\sigma'}^{pm\sigma}(x_{s},y_{s})$ as a whole. Furthermore we
arrange the scattering amplitudes $\phi _{p'm'\sigma ^{\prime
}}^{pm\sigma }$ into another column vector $\phi$ of length $2M$ (
$M=\sum_{p}N_{p}$ is the total number of transverse sites on all
leads). Of course, these two vectors are the central quantities in
our scattering problem and we want to solve them from the
Schrodinger equations. As different from the continuum model, the
Schrodinger equations equations has a lattice form( in real space,
there is \emph{one separate equation} centered on each site and
spin index ) in our case and the normal boundary conditions has
different appearance. Form Eq.(\ref{eq:swavefunction}) is the
general linear combination of scattering eigen-states in the
leads, thus for all sites on the leads except the \emph{first row}
sites, the Schrodinger equation is satisfied automatically. The
Schrodinger equation for \emph{first row} sites, however, has a
different form due to its coupling with the system. On the other
hand, all sites in the system has normal form of Schrodinger
equation for a closed system \emph{except} the boundary sites
which has coupling to the \emph{first row} sites in the leads.
These two \emph{connecting} conditions constitute the boundary
condition in our problem. To be explicit,we write down Schrodinger
equations centered on each site in the system and on \emph{first
row} of all leads separately as:
\begin{widetext}
\begin{eqnarray}
&E\psi
_{\sigma'}^{pm\sigma}(y'_s)&=\sum_{y''_s\sigma''}H_{sys}(y'_s\sigma',y''_s\sigma'')\psi
_{\sigma''}^{pm\sigma}(y''_s)-\sum_{p',y}t_{p's}\delta_{y'_s,n_{p'y}}
\psi _{\sigma'}^{pm\sigma}(1,y_{p'}) \nonumber\\
&E\psi _{\sigma'}^{pm\sigma}(1,y'_{p'})& =
\sum_{y''}H_{leads}(y'_{p'}\sigma',y''_{p'}\sigma')\psi
_{\sigma'}^{pm\sigma}(y''_{p'})-\sum_{y_s}t_{p's}\delta_{y_s,n_{p'y'}}
\psi _{\sigma'}^{pm\sigma}(y_s) \label{eq:boundary}
\end{eqnarray}
\end{widetext}
Note on the above equation we have used $y'_s,y''_s$ to denote a
site in the system and $y'_{p'},y''_{p'}$ to denote a site in lead
$p'$. However,at the same time we also used double index
$(1,y'_{p'})$ etc. for first row sites in lead $p'$ for clarity. $%
n_{p'y'}$ denotes the boundary site in the system near to \emph{first row} site $%
(1,y'_{p'})$ of lead $p'$. Substituting
Eq.(\ref{eq:swavefunction}) into Eq.(\ref{eq:boundary}) and after
some routine algebra, we arrive to the following matrix equations
for the unknown vectors $\psi _{s}$ and $\phi$:

\begin{subequations}
\label{eq:eq1eq2}
\begin{equation}
\textbf{A}\psi _{s}=\textbf{b}+\textbf{B}\phi \label{eq:eq1}
\end{equation}
\begin{equation}
\textbf{C}\phi =\textbf{d}+\textbf{D}\psi _{s}
 \label{eq:eq2}
 \end{equation}
\end{subequations}

Here $\textbf{A},\textbf{B},\textbf{C},\textbf{D}$ are matrices
with dimensions $2N\times 2N,2N\times 2M,2M\times 2M,2M\times 2N$
respectively. They are determined by the Hamiltonian and the
geometric structure of the entire system. $\textbf{b},\textbf{d}$
are constant vectors determined by incident wave. These equations
reflect the mutual influence of the leads and system through
interface hopping. With some algebra, we can write down the matrix
elements explicitly as following:

\begin{eqnarray}
 \textbf{A}&=&E-H_{sys}\nonumber\\
\textbf{B}(n_{p''y}\sigma ^{\prime \prime },p^{\prime }m^{\prime
}\sigma ^{\prime })&=&-\delta _{p^{\prime \prime },p^{\prime
}}\delta _{\sigma ^{\prime \prime },\sigma ^{\prime }}t_{p^{\prime
}s}\chi _{m'}^{p'}(y_{p^{\prime
\prime }})e^{ik_{m'}^{p'}} \nonumber\\
\textbf{D}(p'm'\sigma',n_{p''y}\sigma'')&=&-\delta _{p^{\prime
\prime },p^{\prime }}\delta _{\sigma ^{\prime \prime },\sigma
^{\prime }}t_{p^{\prime }s}\chi _{m'}^{p'}(y_{p''}) \nonumber\\
\textbf{C}(p^{\prime }m^{\prime }\sigma ^{\prime },p^{\prime
\prime }m^{\prime \prime }\sigma ^{\prime \prime })&=&-\delta
_{p^{\prime \prime },p^{\prime }}\delta _{\sigma ^{\prime \prime
},\sigma ^{\prime }}\delta _{m^{\prime \prime
}m^{\prime }}t_{p^{\prime }}\nonumber\\
\textbf{ b}(n_{p'y}\sigma')&=&-\delta _{pp^{\prime
}}\delta _{\sigma \sigma ^{\prime }}t_{ps}\chi _{m}^{p}(y_{p'})e^{-ik_{m}^{p}}\nonumber\\
\textbf{d}(p^{\prime }m^{\prime }\sigma ^{\prime })&=&\delta
_{pp^{\prime }}\delta _{mm^{\prime }}\delta _{\sigma \sigma
^{\prime }}t_{p} \label{eq:matrix}
\end{eqnarray}

The indices for lead, transverse mode, site and spin can take all
possible values. All other matrix elements not included in
Eq.(\ref{eq:matrix}) are zero. Note for some corner points in the
SO system, there may be nearest neighbor points in different
leads. These corner sites have multiple notation in the above
scheme. However, it causes no problem because in real programming,
of course, we will assign certain index to one point.

Note matrix $\textbf{C}$ is diagonal and can be simply inverted.
Defining matrix $\Sigma^R=\textbf{B}\textbf{C}^{-1}\textbf{D}$, it
is easy to see that its elements are:
\begin{equation}
\Sigma^R(n_{y_{1p'}}\sigma',n_{y_{2p'}}\sigma')=-\sum\limits_{m'}\frac{t_{p's}^{2}}{t_{p'}}\chi
_{m'}^{p'}(y_{1})   \chi _{m'}^{p'}(y_{2})e^{ik_{m'}^{p'}}
\label{eq:selfenergy}
\end{equation}
and all other elements are zero. From Eq.(\ref{eq:eq1eq2}) we
have:
\begin{equation}
\psi _{s} =
(\textbf{A}-\Sigma^R)^{-1}(\textbf{b}+\textbf{B}\textbf{C}^{-1}\textbf{d})=G^{R}\textbf{g}
\label{eq:greenr}
\end{equation}

$G^{R}=(\textbf{A}-\Sigma^R)^{-1}$ is, by definition,the retarded
Green's function of the system under the influence of existence of
leads. In Eq.(\ref{eq:greenr}) we defined the \emph{effective}
incident wave for the boundary sites,
$\textbf{g}=\textbf{b}+\textbf{B}\textbf{C}^{-1}\textbf{d}$ whose
nonzero elements are:
\begin{eqnarray}  \textbf{g}(n_{y_{p'}}\sigma ^{\prime })=2i\delta
_{pp'}\delta _{\sigma \sigma'}t_{ps}\sin (k_{m}^p)\chi _{m}^p(y)
\end{eqnarray}

from Eq.(\ref{eq:eq2}),we have
$\phi=\textbf{C}^{-1}\textbf{d}+\textbf{C}^{-1}\textbf{D}G^{R}\textbf{g}$.
By expanding this expression we obtain explicitly the scattering
amplitudes:
\begin{widetext}
\begin{eqnarray}
\phi _{p'm'\sigma'}^{pm\sigma}=-\delta _{pp^{\prime }}\delta
_{mm'}\delta _{\sigma
\sigma'}+t_{p'}^{-1}t_{p's}\sum_{y_{p},y_{p'}}\chi
_{m'}^{p'}(y_{p'})G^{R}(n_{y_{p'}}\sigma ',n_{y_{p}}\sigma
)[2it_{ps}\sin (k_{m}^{p})\chi _{m}(y_{p})] \label{feiexpresssion}
\end{eqnarray}
\end{widetext}

\section{General discussion of rigorous property}
\subsection{The Green's function formula}
Now let's calculate the multi-terminal transmission probability in
the Landauer-Buttiker theory\cite{data}, $T_{p'\sigma'
}^{p\sigma}=\sum\limits_{m,m^{\prime }}\left\vert \phi
_{p'm'\sigma' }^{pm\sigma}\right\vert
^{2}\frac{v_{p'm'}}{v_{pm}}$, where $v_{p'm'}=\frac{1}{\hbar}2t_{p'}\sin (k_{m'}^{p'})$ is the velocity of mode $m'$ in lead $p'$. With Eq.(\ref{feiexpresssion}), for $p\neq p^{\prime }$,we get:%

\begin{eqnarray}
T_{p'\sigma'}^{p\sigma}&=&\sum\limits_{m,m'}\left\vert\phi
_{p'm'\sigma'}^{pm\sigma}\right\vert
^{2}\frac{v_{p'm'}}{v_{pm}}\nonumber\\
&=&Tr(\Gamma ^{p}G_{\sigma \sigma ^{\prime }}^{A}\Gamma
^{p^{\prime }}G_{\sigma ^{\prime }\sigma }^{R})
\end{eqnarray}

\bigskip where%
\[
\Gamma ^{p}(y_{p},\overline{y}_{p})=\sum\limits_{m}(\frac{t_{ps}}{t_{p}}%
)^{2}\chi _{m}(y_{p})v_{pm}\chi _{m}(\overline{y}_{p})
\]

and $G^R_{\sigma\sigma'}$ and $G^A_{\sigma\sigma'}$ are
spin-resolved retarded and advanced Green's functions. This is
just the most frequently used Green's function formula\cite{data}.
Here we have deduced it in a rigorous manner from the scattering
wave description. We point out here that self energy $\Sigma^R$
deduced in many text books\cite{data} have neglected contribution
from evanescent modes, while our expression in
Eq.(\ref{eq:selfenergy}) with longitudinal wave vector given by
Eq.(\ref{eq:wave vector}) was rigorous.
\bigskip

\subsection{Some rigorous property of equilibrium state}
\subsubsection{flowing/partial and full spectral function}
 Now, following Landauer's spirit\cite{data}, we assume the reservoirs
connecting the leads at infinity will feed one-way moving
particles to leads according to their own chemical potential.
Let's normalize the scattering wave function $\psi^{pm\sigma}$ so
that there is one particle in the incident wave, i.e., we change
$e^{ik_{m}^p\sigma}$ to $e^{ik_{m}^p\sigma}/\sqrt{L}$,where
$L\rightarrow\infty$ is the length of lead $p$. Meanwhile, the
density of states in lead $p$ is
$\frac{L}{2\pi}\frac{dk}{dE}=\frac{L}{2\pi \hbar v_{pm}}$. Now
let's we add all incident channel $(m\sigma )$ of lead $p$ with
corresponding density of states(DOS). Then, we get the partial(or
flowing) density of states arising from incident waves in lead $p$
as following:
\begin{eqnarray}
&\frac{1}{2\pi}&\sum\limits_{m\sigma }\psi _{s}^{pm\sigma\ast
}(y^{\prime }\sigma ^{\prime })\psi _{s}^{pm\sigma}(y^{\prime
}\sigma ^{\prime })1/\hbar v_{pm}\nonumber\\
&=\frac{1}{2\pi}&\sum\limits_{y_{1},y_{2},\sigma }G^{R}(y^{\prime
}\sigma ^{\prime },n_{py_{1}}\sigma )\Gamma
^{p}(y_{1},y_{2})G^{A}(n_{py_{2}}\sigma ,y^{\prime }\sigma
^{\prime })\nonumber\\
&=\frac{1}{2\pi}&A^{p}(y^{\prime }\sigma ^{\prime },y^{\prime
}\sigma ^{\prime })
\end{eqnarray}
where we defined $A^{p}(y^{\prime }\sigma ^{\prime },y^{\prime
}\sigma ^{\prime })$ as the partial/flowing spectral function
arising from current incidence in lead $p$.

Consider the equilibrium state when all reservoirs has the same
chemical potential. Then, at any energy, the carriers is incident
equally from all reservoirs and we should sum up the
incident lead index $p$ for scattering wave functions when calculating flowing density of states:%
\begin{eqnarray}
&\frac{1}{2\pi}&\sum\limits_{p,m\sigma }\psi
_{s}^{pm\sigma\ast}(y'\sigma ')\psi
_{s}^{p,m\sigma}(y'\sigma')1/\hbar v_{pm}\nonumber\\
&=\frac{1}{2\pi}&\sum\limits_{y_{1},y_{2}}G^{R}(y'\sigma',n_{y_{1}}\sigma
)\Gamma (y_{1},y_{2})G^{A}(n_{y_{2}}\sigma,y'\sigma')
\end{eqnarray}
$n_{y_{1}},n_{y_{2}}$ refer to all boundary sites in the system
which has nearest hopping term to leads, and
$\Gamma=\sum\limits_{p}\Gamma ^{p}$. Furthermore, we can prove an
important relation\cite{data}. Since
\begin{equation*}
\left[ G^{R}\right] ^{-1}-%
\left[ G^{A}\right]
^{-1}=\Sigma^{A}-\Sigma^{R}=i\sum\limits_{p}\Gamma ^{p}=i\Gamma
\end{equation*}
then,
\begin{equation*}
G^{A}-G^{R}=iG^{R}\Gamma G^{A}\equiv iA
\end{equation*}
where $A =i(G^{R}-G^{A})$ is the full spectral function which
plays a role of a generalized density of states inside the SO
system taking the presence of all leads into account. In an open
system, it's not absolute clear how to define state and density of
states. Here through our deduction we have made explicit the
physical meaning of $A $ from the wave function description. We
emphasize the particular boundary condition used in the leads,
i.e., there is a thermal reservoir with the same chemical
potential $\mu $, in the far end for every lead, in the
Landauer-Buttiker sense. For non-equilibrium state, i.e., not all
of the leads have same chemical potential and there will be net
current in the leads and system. At that time,we are interested in
flowing density $A^{p}$ rather than total density $A$.

On the above we used only the diagonal elements of $A^{p}$ and
$A$. It is convenient to define the non-diagonal elements as
flowing spin density density matrix:

\begin{eqnarray}
&A^{p}(y'\alpha,&y''\beta)=\sum\limits_{m\sigma}\psi
_{s}^{\ast}(y''\beta )\psi _{s}(y'\alpha
)1/\hbar v_{pm}\nonumber\\
&=&\sum\limits_{y_{1},y_{2},\sigma }G^{R}(y'\alpha
,n_{y_{1p}}\sigma )\Gamma ^{p}(y_{1},y_{2})G^{A}(n_{y_{2p}}\sigma
,y''\beta )\nonumber\\
&A(y'\alpha,&y''\beta)=\sum\limits_{p,m\sigma }\psi _{s}^{\ast
}(y^{\prime \prime }\beta )\psi _{s}(y^{\prime }\alpha
)1/\hbar v_{pm}\nonumber\\
&=&\sum\limits_{p,y_{1},y_{2},\sigma }G^{R}(y^{\prime }\alpha
,n_{y_{1p}}\sigma )\Gamma (y_{1},y_{2})G^{A}(n_{y_{2p}}\sigma
,y^{\prime \prime }\beta )\nonumber\\
&=&i(G^{R}(y'\alpha,y''\beta)-G^{A}(y'\alpha,y''\beta))
\end{eqnarray}

\subsubsection{All $T$ odd quantities vanishes in time reversal symmetric system}
  Now let's prove that in the Landauer-Buttiker
description, for a time-reversal symmetric Hamiltonian, all
$T$-odd quantities in equilibrium are zero:

Consider a local quantity:
\begin{equation}
\hat{O}^{ij}=\sum_{\alpha\beta}(C_{i\alpha}^{\dagger}O_{\alpha\beta}^{ij}C_{j\beta}+C_{j\alpha}^{\dagger}O_{\alpha\beta}^{ji}C_{i\beta})
\label{eq:physicalquantity}
\end{equation}

Typical examples: (1)for spin density on site i, we choose
$O^{ij}=\frac{\hbar}{4}\delta_{ij}\vec{\sigma}$. (2)for normally
defined charge current on link $(i,j)$, we set
$O^{ij}=-O^{ji}=ieI_{2\times2}$ where $I_{2\times2}$ is the 2d
unit matrix, (3)for normally defined spin current on link $(i,j)$,
we have $O^{ij}=-O^{ji}=\frac{\hbar}{2}i\vec{\sigma}$. In general
we require $O^{ij}=O^{ji\dagger}$ for physical quantity.

Under time reversal manipulation, $\emph{T}=i\sigma_{y}K$, where
$K$ is the conjugate operator,  $\hat{O}^{ij}$ will transform as
$O^{ij}_{\alpha\beta}\rightarrow
(-1)^{\alpha+\beta}O_{\bar{\alpha}\bar{\beta}}^{ij*}$. There are
two classes of physical quantities, depending on their symmetry
under time reversal operation, i.e., either $T$-even or $T$-odd
quantities. Spin density and normally defined link currents are
$T$-odd quantities. In contrast, the normally defined spin
currents are $T$-even quantity.

  Now, let's calculate the expectation value of $\hat{O}^{ij}$ in
mesoscopic equilibrium state. By definition, this expectation
value per energy interval is:
\begin{eqnarray}
\langle\hat{O}^{ij}\rangle&=&\frac{1}{2\pi}\sum\limits_{pm\sigma
}\psi _{s}^{pm\sigma\ast}(i\alpha)O_{\alpha\beta}^{ij}\psi
_{s}^{pm\sigma}(j\beta)1/\hbar v_{pm}+h.c.\nonumber\\
&=&\frac{1}{2\pi}\sum_{\alpha\beta}[A(j\beta,i\alpha)O_{\alpha\beta}^{ij}+h.c.]\nonumber\\
&=&\frac{1}{2\pi}i\sum_{\alpha\beta}[(G^R-G^A)_{j\beta,i\alpha}O_{\alpha\beta}^{ij}+(G^R-G^A)_{i\alpha,j\beta}O_{\beta\alpha}^{ji}]\nonumber\\
\label{eq:21}
\end{eqnarray}
where we have used the relation $(G^R)\dag=G^A$. Moreover,the
matrices $(G^R)$ and $G^A$ are related by time reversal operation.
Generally, by definition, we have:
\begin{equation*}
(TG^R_{B}T^{-1})_{i\alpha,j\beta}=(-1)^{\alpha+\beta}G_{i\bar{\alpha},j\bar{\beta}}^{R*}(B)=G_{j\beta,i\alpha}^{A}(-B)
\end{equation*}
Here, for generality of discussion, we have used magnetic field B
to break $T$ symmetry explicitly. Furthermore, let's suppose that
the Hamiltonian is $T$ invariant, then
\begin{equation}
(TG^RT^{-1})_{i\alpha,j\beta}=(-1)^{\alpha+\beta}G_{i\bar{\alpha},j\bar{\beta}}^{R*}=G_{j\beta,i\alpha}^{A}
\end{equation}

The four terms in Eq.(\ref{eq:21}) can be grouped into two pairs,

\begin{widetext}
\begin{eqnarray*}
\sum_{\alpha\beta}[G^R_{j\beta,i\alpha}O_{\alpha\beta}^{ij}-G^A_{i\alpha,j\beta}O_{\beta\alpha}^{ji}]
+\sum_{\alpha\beta}[-G^A_{j\beta,i\alpha}O_{\alpha\beta}^{ij}+G^R_{i\alpha,j\beta}O_{\beta\alpha}^{ji}]\nonumber\\
=\sum_{\alpha\beta}[G^R_{j\beta,i\alpha}O_{\alpha\beta}^{ij}-(-1)^{\alpha+\beta}G^{R*}_{j\bar{\beta},i\bar{\alpha}}O_{\alpha\beta}^{ij*}]
+\sum_{\alpha\beta}[-(-1)^{\alpha+\beta}G^{R*}_{i\bar{\alpha},j\bar{\beta}}O_{\beta\alpha}^{ji*}+G^R_{i\alpha,j\beta}O_{\beta\alpha}^{ji}]
\end{eqnarray*}
\end{widetext}
Now, suppose $\hat{O}$ is $T$ odd, then,
$(-1)^{\alpha+\beta}O_{\alpha\beta}^{ij*}=-O_{\bar{\alpha}\bar{\beta}}^{ij}$
. Thus we have
\begin{widetext}
\begin{eqnarray}
\sum_{\alpha\beta}[G^R_{j\beta,i\alpha}O_{\alpha\beta}^{ij}-G^A_{i\alpha,j\beta}O_{\beta\alpha}^{ji}]+
\sum_{\alpha\beta}[-G^A_{j\beta,i\alpha}O_{\alpha\beta}^{ij}+G^R_{i\alpha,j\beta}O_{\beta\alpha}^{ji}]\nonumber\\
=\sum_{\alpha\beta}[G^R_{j\beta,i\alpha}O_{\alpha\beta}^{ij}+G^{R*}_{j\bar{\beta},i\bar{\alpha}}O_{\bar{\alpha}\bar{\beta}}^{ij*}]+
\sum_{\alpha\beta}[G^{R*}_{i\bar{\alpha},j\bar{\beta}}O_{\bar{\beta}\bar{\alpha}}^{ji*}+G^R_{i\alpha,j\beta}O_{\beta\alpha}^{ji}]
\label{eq:real}
\end{eqnarray}
\end{widetext}
evidently, each summation on the right hand side of
Eq.(\ref{eq:real}) is real. So, the average value for
\emph{physical} quantity in Eq.(\ref{eq:21}) should be zero.

 On the above we have proved that for any time reversal symmetric system, all $T$-odd quantities, such as
spin density and local charge current, vanish in equilibrium
state. This result seems to be a transparent truth. However, up to
our knowledge, though many people accept it to be true, no one has
proved it rigorously in Landauer's mesoscopic structure. The
corresponding fact for a closed system in equilibrium is trivial,
because in closed system,  each quantum state has its $T$ partner
state(except the $T$ invariant states for which the average of
$T$-odd operator should be zero), thus for a $T$-odd operator, the
sum of average is zero. However, in our open system case, for
equilibrium state, we have an ensemble of scattering wave
functions incident from all leads at fermi energy while all the
states are not manifestly $T$ paired.

Under finite terminal bias, the system is driven into
non-equilibrium state. Due to above property, the value of $T$-odd
quantity is determined by fermi surface property for small bias.
This is in accordance with fermi liquid theory. But for $T$-even
quantities, e.g., spin current, we cannot reach to such a
conclusion. In fact, we need other symmetry considerations to
interpret some $T$-even quantities to be fermi surface
property\cite{NikolicPRB2006}.

\begin{figure}[tbh]
\includegraphics[width=6cm,height=2cm]{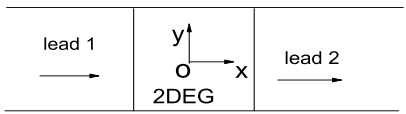}
\caption{schematic geometry of two-terminal structure}
\label{fig:2terminal}
\end{figure}

For later use, let's consider a geometrically symmetric system
with double terminals, as shown in Fig.\ref{fig:2terminal}. Since
the spin density vanishes in equilibrium state, according to
Eq.(\ref{eq:21})we have:
\begin{equation}
<S_{\alpha}(x,y)>_{eq}=<S_{\alpha}(x,y)>_{I}+<S_{\alpha}(x,y)>_{-I}=0
\label{eq:spinvanish}
\end{equation}
where $<S_{\alpha}(x,y)>_{I}$ means partial/flowing average of
spin density due to scattering states incident from lead 1, and
$<S_{\alpha}(x,y)>_{-I}$ means flowing spin density due to
scattering waves incident from lead 2. This relationship will be
employed to derive an important symmetry properties in next
section.

\bigskip

\subsubsection{equilibrium terminal spin current is always zero}

 Next, for ease of reference, we will give a explicit proof that terminal spin current also vanishes in
 equilibrium state(which is known by others\cite{thank}), irrespective of the $T$ symmetry of the
 Hamiltonian. In linear transport regime, we neglect the energy dependence
 of scattering matrices, then the terminal spin current polarized in $\sigma$ direction
 in $p$'th lead can be calculated as,
\begin{eqnarray}
I_{p}^{\sigma }=-\frac{e}{4\pi
}\{\sum_{q\sigma'}^{\prime}[(T_{p\sigma
}^{q\sigma'}-T_{p\overline{ \sigma
}}^{q\sigma'})V_{q}-(T_{q\sigma'}^{p\sigma }-T_{q\sigma'}^{p\overline{\sigma }%
})V_{p}]\nonumber\\
+2(R_{p\sigma}^{p\overline{\sigma}}-R_{p\overline{\sigma
}}^{p\sigma})V_{p}\}
\end{eqnarray}

Where we used symbol $R$ to represent reflection probability in
lead $p$, i.e., we define
$R_{p\sigma}^{p\overline{\sigma}}=T_{p\sigma}^{p\overline{\sigma}}$
etc for clarity. On the above equation, the summation is taken
over all terminals except $p$. The second term in
\verb"{"$\cdots$\verb"}" describes contribution to spin current
due to injection and reflection wave in $p$'th lead. This term is
zero for normal current, where summation of all spin indices is
performed.

For equilibrium state, we have $V_{q}=$\textsl{constant} for all
$q$. Let's firstly write down the following identities:
\begin{eqnarray}
-\sum\limits_{q\sigma ^{\prime }}^{\prime}T_{q\sigma ^{\prime
}}^{p\sigma }-R_{p\overline{\sigma }}^{p\sigma }-R_{p\sigma
}^{p\sigma
}=-M_{p}\nonumber\\
\sum\limits_{q\sigma ^{\prime }}^{\prime}T_{q\sigma
^{\prime }}^{p\overline{\sigma }}+R_{p\sigma }^{p\overline{\sigma }}%
+R_{p\overline{\sigma }}^{p\overline{\sigma }}=M_{p}\nonumber\\
\sum\limits_{q\sigma ^{\prime }}^{\prime}T_{p\sigma }^{q\sigma
^{\prime }}+R_{p\sigma
}^{p\overline{\sigma }}+R_{p\sigma }^{p\sigma }=M_{p}\nonumber\\
-\sum\limits_{q\sigma ^{\prime }}^{\prime}T_{p%
\overline{\sigma }}^{q\sigma ^{\prime }}-R_{p\overline{\sigma }}^{p\overline{%
\sigma }}-R_{p\overline{\sigma }}^{p\sigma }=-M_{p}\nonumber\\
\end{eqnarray}

Here $M_{p}$ denotes the number of conducting mode in lead $p$.
 The first two equations are simple sum rules. They describe current
conservation: one incident electron in lead $p$ must go somewhere.
The last two equations follows from the unitarity of scattering
matrix. Or,we can deduce them from the first two since
$T_{p\sigma}^{q\sigma'}(B)=T_{q\overline{\sigma
}'}^{p\overline{\sigma
}}(-B)$,$R_{p\sigma}^{p\sigma'}(B)=R_{p\sigma'}^{p\sigma}(-B)$ due
to $T$ transformation property of scattering amplitudes. Adding up
all the above equations we get $I_{p(eq)}^{\sigma }=0$.

Thus, we have showed explicitly that terminal equilibrium spin
currents are identically zero, irrespective of details of
Hamiltonian of the system(even in the presence of magnetic field)
and geometry structure. This result is parallel to the fact that
terminal equilibrium charge current always vanish, while current
density may be nonzero inside system when the system is placed
under magnetic field.

E.I.Rashba pointed out\cite{RashbaPRB2003} that in a bulk 2DEG
system with pure Rashba type of SO coupling, in-plane polarized
spin current may have non-zero value in equilibrium. This simple
result raised the problem of how to define spin current properly
in SO systems. A much concerned problem is that whether such
background spin current useful? From the \emph{practical} point of
view, the useful quantity is terminal spin current\cite{pareek}.
In a conceived all-electric devise, we should add nonmagnetic
contact to induct spin current out of SO system. However,
according to the above rigorous result, the resulting equilibrium
spin current in added terminals around the SO system should always
vanish and we cannot simply make use of it. How does spin current
decay abruptly at the boundary between lead and bulk system is an
interesting problem to be clarified. Such an understanding may
help to bridge the apparent gap between mesoscopic approaches and
macroscopic approaches\cite{Jsinova2005}(say, linear response
approach).

 When there is current flowing through the SO system, there will be induced
spin current on attached leads. In the following we will discuss
some symmetry property of such spin currents in a two terminal
structure. We will reveal an \emph{emergent} continuity property
of spin currents due to symmetry of geometry and SO Hamiltonian.

For later use,  here we write down the following equation for a
two terminal system:
  \begin{equation}
I_{2(eq)}^{\sigma}=I_{2(I)}^{\sigma }+I_{2(-I)}^{\sigma }=0
\label{eq:spincurrentvanish}
\end{equation}
where $I_{2(eq)}^{\sigma}$ is the equilibrium spin current in lead
2, and  $I_{2(I)}^{\sigma }$ and $I_{2(-I)}^{\sigma }$ denote spin
current in lead 2 under the condition of fixed charge current I
(flowing from lead $1\rightarrow 2$) and -I (flowing from lead
$2\rightarrow 1$) respectively.

\section{Some symmetry property of typical Spin-orbit coupling systems}
Now let's discuss the symmetry of typical Spin-orbit coupling
system in which a SO bar is symmetrically attached by two ideal
leads. Let's model the SO bar by a Rashba Hamiltonian,
$H_{so}=\alpha (\hat{\sigma}\times \vec{k})\cdot \vec{z}$, whose
discrete version is just Eq.(\ref{eq:rashba}). We assume a
rectangular geometry as shown in Fig 2. Firstly, let's analyze
symmetry properties of the model Hamiltonian. For pure system,
it's easy to see from Fig.2 and Eq.(\ref{eq:1}),
Eq.(\ref{eq:rashba}), that the Hamiltonian has combined symmetry
of real space reflection and spin space rotation:
$\sigma_{y}P_{x}$ and $\sigma_{x}P_{y}$, where $P_{x}$ and $P_{y}$
denotes real space reflection $y\Rightarrow -y$ and $x\Rightarrow
-x$, respectively, $\sigma_{x}$ and $\sigma_{y}$ acting on spinor
wave function on every site are spin rotational manipulation
around $x$ axis and $y$ axis, respectively.  Moreover, when there
is no external magnetic field, the Hamiltonian for the entire
system is $T$ symmetric. From these symmetries, we can obtain the
following symmetry relations:
\begin{eqnarray}
<S_{x,z}(x,y)> _{I}&=&-<S_{x,z}(x,-y)> _{I} \nonumber\\
<S_{y}(x,y)> _{I}&=&< S_{y}(x,-y)> _{I}\nonumber\\ < S_{x}(x,y)>
_{I}&=&-<S_{x}(-x,y)> _{I} \nonumber\\ < S_{y,z}(x,y)>
_{I}&=&<S_{y,z}(-x,y)> _{I} \label{eq:symmetryR}
\end{eqnarray}
$<\cdots> _{I}$ means taking ensemble average for given
longitudinal current $I$ incident from lead 1. The first two lines
are results of symmetry $\sigma_{y}P_{x}$ which has been known
widely before\cite{seeSQShen2004}. The second two, however, are
results from $\sigma_{x}P_{y}$ and $T$ symmetry together. From
$\sigma_{x}P_{y}$, we get:$<S_{y,z}(x,y)> _{I}=-<S_{y,z}(-x,y)>
_{-I},<S_{x}(x,y)> _{I}=< S_{x}(-x,y)> _{-I}$. From time reversal
symmetry,we have Eq.(\ref{eq:spinvanish}). Combining these two, we
obtain the last two lines in Eq.(\ref{eq:symmetryR}).

 Similar symmetry relations can be easily obtained for Dresselhaus
 SO coupling model\cite{SQShen2004} for which the discrete Hamiltonian reads:
\begin{equation}
H_{so}^D=t^{D}\sum_{R_{i}}[i(\Psi _{R_{i}}^{\dag }\sigma ^{y}\Psi
_{R_{i}+y}-\Psi _{R_{i}}^{\dag }\sigma ^{x}\Psi _{R_{i}+x})+h.c.]
\label{Dresselhaus}
\end{equation}

As can be easily seen, we have  symmetry
 $\sigma_{x}P_{x}$, $\sigma_{y}P_{y}$ and $T$ in this Dresselhaus
 model. In this case, we get the following symmetry relations:
\begin{eqnarray}
<S_{y,z}(x,y)> _{I}&=&-<S_{y,z}(x,-y)> _{I} \nonumber\\
<S_{x}(x,y)> _{I}&=&< S_{x}(x,-y)> _{I}\nonumber\\ < S_{y}(x,y)>
_{I}&=&-<S_{y}(-x,y)> _{I} \nonumber\\ < S_{x,z}(x,y)>
_{I}&=&<S_{x,z}(-x,y)> _{I} \label{eq:symmetryD}
\end{eqnarray}

In the presence of both $H_{so}^R$ and $H_{so}^D$ with geometry
shown in Fig.2, it can be checked easily that the total
Hamiltonian is invariant under $\sigma_{z}P_{x}P_{y}$. Thus we
have symmetry relations:
\begin{eqnarray}
<S_{z}(x,y)> _{I}&=&-<S_{z}(-x,-y)> _{I}\nonumber\\
<S_{x,y}(x,y)> _{I}&=&< S_{x,y}(-x,-y)> _{I}\nonumber\\
\label{eq:symmetryRD}
\end{eqnarray}

Up to our knowledge, not all of the symmetry relations
Eq.(\ref{eq:symmetryR}), Eq.({\ref{eq:symmetryD}) and
Eq.({\ref{eq:symmetryRD}) have  been reported previously. In
particular, Eq.({\ref{eq:symmetryRD}) and the last two lines in
Eq.(\ref{eq:symmetryR}) as well as in Eq.({\ref{eq:symmetryD})
follow from $T$ symmetry together with geometric symmetry( spacial
reflection combining a spin rotation manipulation) are firstly
found in this paper. These relations provide important accuracy
 test for experimental measurements and numerical calculations.
For example, for a two-terminal structure with Rashba coupling in
the middle SO bar, the theoretical assertion made in
\cite{NikolicPRL2005} that longitudinal spin component is
insensitive to the reversal of the bias voltage is contrary to our
Eq.(\ref{eq:spinvanish}), the numerical
result\cite{NikolicPRL2005} that spin density $<S_{z}(x,y)>_V$ is
\emph{not} symmetric with $x\rightarrow{-x}$ is contrary to the
last line of Eq.(\ref{eq:symmetryR}).

The issue of spin current has triggered hot discussion recently.
The central problem lies in that spin is non-conservative in
spin-orbit coupling system. In bulk systems, proper definition and
physical consequence of spin current are currently under debate.
However, in a multi-terminal structure, spin current on the leads
can be defined unambiguously. In the following let's analyze the
symmetry of spin current in two terminal structure with geometric
symmetry( spacial reflection combining a spin rotation
manipulation). The terminal spin current on a particular lead was
a summation of link spin current(see discussion below
Eq.(\ref{eq:physicalquantity})) in transverse direction. Then, for
a system with both $H_{so}^R$ and $H_{so}^D$ present, by taking
into account Eq.(\ref{eq:spincurrentvanish}) together with
geometry symmetry and following similar argument in the derivation
of  symmetry relations for spin density, we get a symmetry
relation similar to Eq.(\ref{eq:symmetryRD}):

\begin{eqnarray}
&I_{2(I)}^{z}&=-I_{1(I)}^{z}\nonumber\\
&I_{2(I)}^{x,y}&=I_{1(I)}^{x,y} \label{eq:symmetrySC}
\end{eqnarray}

Such symmetry of terminal spin current is directly related to
symmetry of spin polarization Eq.(\ref{eq:symmetryRD}) since the
terminal spin current can be imagined to be injected from boundary
spin density. The symmetry connection between these two quantities
is clearly depicted in Fig.(\ref{fig:spincur}).
\begin{figure}[tbh]
\includegraphics[width=7cm,height=7cm]{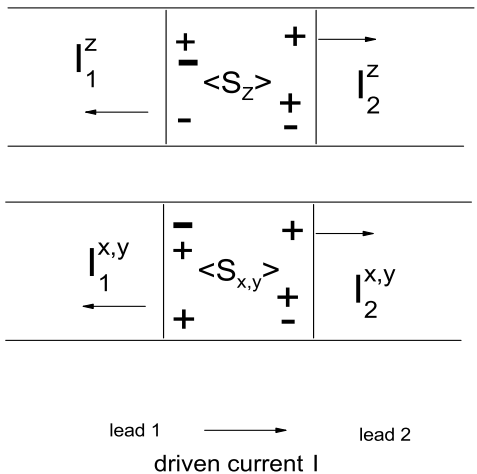}
\caption{schematic picture of symmetric relation between boundary
spin density and terminal spin current in a two-terminal structure
} \label{fig:spincur}
\end{figure}

 Interestingly,
$I_{2(I)}^{z}$ can be interpreted as \emph{conserved} current
because the quantity of it injecting into and leaving out of the
SO region is identical. This \emph{emergent} continuity property,
however, is the result of  geometric symmetry. In contrast to a
recently proposed spin current definition\cite{shijunren} in bulk
system,
 for which a coarse graining process(though physically beautiful, such
process is hard to done rigorously) is necessary, the continuity
of spin current in our case is due to dynamical and geometric
symmetry. This interpretation has at least the advantage of
physical transparency.
  The different property of terminal spin current polarized normal and parallel
to the 2DEG plane is an interesting topic to discuss. As mentioned
on the above, the equilibrium in-plane spin current is not useable
in an all-electric device. Under the condition of flowing current
in an ideal system, terminal in-plane spin current lacks a simple
physical interpretation as a current while the out-of-plane spin
current does. From these simple symmetry relations, we believe
these components may play very different role in spin
transportation in real systems. It is interesting to consider more
geometries to further expose such symmetry differences and study
the different role of in-plane and out-of-plane spin current to
spin transportation in real dissipative systems.

\begin{figure}[tbh]
\includegraphics[width=7cm,height=7cm]{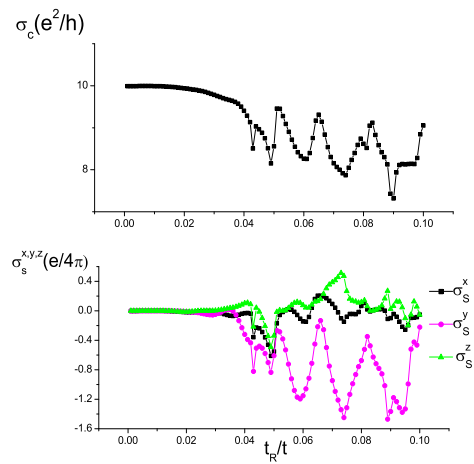}
\caption{The charge and spin conductances as a function of Rashba
coupling constant $t_R$. The system is taken to be $100\times 40$
sized and Dresselhaus coupling constant is fixed to be 0.02t}
\label{fig:conductances}
\end{figure}

  To have a quantitative impression on spin current, we performed some numerical
  calculation. Let's consider a two-terminal system with both Rashba and Dresselhaus coupling. We adopt a
discrete representation and fix $t^D/t=0.02$. The current is
assumed to flow from lead 1 to lead 2 for a system with $100\times
40$ lattice sites and both leads are of the same width as system,
the charge conductance $\sigma_c$ and spin conductances
$\sigma_{s}^{x,y,z}$ are shown in Fig.(\ref{fig:conductances}).
The spin conductances are much smaller compared to charge
conductances and both of them oscillate rapidly in large $t_R/t_D$
regime. For $t_R/t_D<2$, $\sigma_c$ decreases smoothly and
$\sigma_{s}^{x,y,z}$ are very small. When $t_R/t_D>2$, both
$\sigma_c$ and $\sigma_{s}^{x,y,z}$ lines will show series of peak
structures. The line-shape of $\sigma_{s}^{y}$ will follow closely
to that of $\sigma_{c}$. On the other hand, the $\sigma_{s}^{x,z}$
values will decrease in the large $t_R$ regime due to symmetry
property stated below. Once spin current become experimentally
detectable, the peaks structures in the
Fig.(\ref{fig:conductances}) are on the first place to be
verified.

 For pure Rashba system, due to $y\rightarrow
-y$ symmetry(combining spin rotation) we have:

\begin{eqnarray}
&I_{2(I)}^{x,z}&=I_{1(I)}^{x,z}=0 \nonumber\\
&I_{2(I)}^{y}&=I_{1(I)}^{y} \label{eq:symmetrySCR}
\end{eqnarray}

similarly, for pure Dresselhaus system, we have:
\begin{eqnarray}
&I_{2(I)}^{y,z}&=I_{1(I)}^{y,z}=0 \nonumber\\
&I_{2(I)}^{x}&=I_{1(I)}^{x} \label{eq:symmetrySCD}
\end{eqnarray}

The symmetry relations
Eq.(\ref{eq:symmetrySC}),(\ref{eq:symmetrySCR}),(\ref{eq:symmetrySCD})
can hold even in the presence of magnetic field(which keep
geometry symmetry), since Eq.(\ref{eq:spincurrentvanish}) is
independent of $T$ symmetry.

 On the above we have obtained some new symmetry relations relating to spin
polarization as well as terminal spin currents for a two-terminal
structure. Some other symmetries for a two-terminal waveguide in
the presence of magnetic field modulations is obtained in
Ref.\cite{zhaifeng}. All these symmetry relations may find
application in the study of spin transport theory.

\section{conclusion}
To sum up, we have set up a detailed formulation to solve
explicitly the scattering wave functions in a multi-terminal
spin-orbit coupled system. We deduced some analytical properties
in the scattering wave function description of mesoscopic physics
following Landauer's spirit. In particular, we have (1) deduced
rigorously the much used Green's function formula. (2) proved
rigorously that all equilibrium value of $T$-odd quantities vanish
in the open multi-terminal structure. (3) reveal some new symmetry
relations for two-terminal structure when the system has a Rashba
or/and Dresselhaus form of SO coupling. These symmetry relations
may provide accuracy test to experimental measurements and
symmetry restriction to numerical calculations.

\begin{acknowledgments}
Y. J. Jiang was supported by Natural Science Fundation of Zhejiang
province ( Grant No.Y605167 ). L. B. Hu was supported by the
National Science Foundation of China ( Grant No.10474022 ) and the
Natural Science Foundation of Guangdong province ( No.05200534 ).
\end{acknowledgments}


\begin{thebibliography}{99}
\bibitem{data}See, e.g., S.Datta, \textit{Electronic transport in mesoscopic systems} (Cambridge University Press, Cambridge, 1997).
\bibitem{Hirsch1999}J.E.Hirsch,Phys.Rev.Lett,{\bf,83},1834(1999). S.Zhang, Phys. Rev. Lett.{\bf,85},393(2000).
\bibitem{MurakamiScience2003}S.Murakami,N.Nagaosa,and S.C.Zhang,Science{\bf
301},1348(2003).
\bibitem{SinovaPRL2004}J.Sinova,D.Culcer,Q.Niu,N.A.Sinitsyn,T.Jungwirth,and
A.H.MacDonald,Phys.Rev.Lett{\bf 92},126603(2004).

\bibitem{Dassarma2004} I. Zutic, J. Fabian, and S. Sarma, Rev. Mod. Phys
\textbf{76}, 323(2004).

\bibitem{semibook2002} D. Awschalom, D. Loss, and N. Samarth, \textit{%
Semiconductor Spintronics and Quantum Computation} ( Springer,
Berlin, 2002).
\bibitem{JPHu2003}J.P.Hu,B.A.Bernevig,and C.J.Wu,Int.J.Mod.Phys.B{\bf 17},5991\\ (2003).
\bibitem{RashbaPRB2003} E. I. Rashba, Phys. Rev. B \textbf{68},
(R)241315(2003); \textit{ibid}.\textbf{70}, (R)201309(R)(2004).
\bibitem{InouePRB2004} J. Inoue, G. E. W. Bauer, and L. W. Molenkamp, Phys.
Rev. B\textbf{70}, 041303(R)(2004).

\bibitem{MishchenkoPRL2004} E. G. Mishchenko, A. V. Shytov, and B. I.
Halperin, Phys. Rev. Lett. \textbf{93}, 226602(2004).


\bibitem{SQShen2004}S.Q.Shen,Phys.Rev.B{\bf 70},R081311(2004).

\bibitem{NikolicPRL2005} B. K. Nikolic, S. Souma, L. P. Zarbo and J. Sinova,
Phys. Rev. Lett \textbf{95}, 046601(2005).

\bibitem{NikolicPRB2005} B. K. Nikolic, L. P. Zarbo and S. Souma, Phys. Rev.
B \textbf{72}, 075361(2005);
\bibitem{NikolicPRB2006} B. K. Nikolic, L. P. Zarbo and S. Souma, Phys. Rev.
B \textbf{73}, 075303(2006).

\bibitem{LShengPRL2005} L. Sheng, D. N. Sheng, and C. S. Ting, Phys. Rev.
Lett. \textbf{94}, 016602(2005);
\bibitem{JLi2005}J.Li, L.B.Hu,and S.Q.Shen,Phys.Rev.B{\bf
71},241305(R)(2005).
\bibitem{Mlushkov05} A. G. Malshukov, L. Y. Wang, C. S. Chu, and K. A. Chao,
Phys. Rev. Lett. \textbf{95}, 146601(2005).

\bibitem{HalprinPRL} H. A. Engel, B. I. Halperin, and E. I. Rashba, Phys.
Rev. Lett. 95, 166605 (2005).

\bibitem{SarmaPRL} W. K. Tse and S. Das Sarma, Phys. Rev. Lett. 96, 056601
(2006).

\bibitem{Usaj2005} A. Reynoso, G. Usaj, and C. A. Balseiro, cond-mat/0511750.

\bibitem{Sarma06} V. M. Galitski, A. A. Burkov, S. Das Sarma,
cond-mat/0601677.

\bibitem{Ma2004} X. H. Ma, L. B. Hu, R. B. Tao, and S. -Q. Shen, Phys. Rev.
B \textbf{70}, 195343(2004); L. B. Hu, J. Gao, and S. -Q. Shen,
Phys. Rev. B 70, 235323(2004).
\bibitem{shijunren}Phys. Rev. Lett. 96, 076604 (2006)

\bibitem{Jsinova2005} J.Sinova, S.Murakami, S.Q.Shen, M.S.Choi, cond-mat/0512054.
\bibitem{Zyang2005}J.Yao and Z.Yang,Phys.Rev.B {\bf 73},033314(2006).
\bibitem{YJJiang2005} Y.J.Jiang, cond-mat/0510664.  Y.J.Jiang, L.B.Hu, cond-mat/0603755.
\bibitem{KatoScience2004} Y. K. Kato, R. C. Myers, A. C. Gossard, and D. D.
Awschalom, Science \textbf{306}, 5703(2004).

\bibitem{WunderlichPRL2005} J. Wunderlich, B. Kaestner, J. Sinova, and T.
Jungwirth, Phys. Rev. Lett. \textbf{94}, 047204(2005).
\bibitem{thank} This point is firstly noticed by B. K. Nikolic et
al. in their paper\cite{NikolicPRB2005}.  Y. J. Jiang owe
gratitude to B. K. Nikolic for informing him this point.
\bibitem{seeSQShen2004}See,.e.g., in
\cite{SQShen2004,NikolicPRL2005,Zyang2005}.
\bibitem{pareek}T.P.Pareek, Phys. Rev. Lett. 92, 076601
(2004).

\bibitem{zhaifeng}F.Zhai and H.Q.Xu, Phys. Rev. Lett. 94, 246601(2005);
\end{thebibliography}
\end{document}